\documentclass[prd,superscriptaddress,nofootinbib,amsmath,amssymb,aps,11pt]{revtex4-2}

\usepackage{bm}
\usepackage{amsfonts}
\usepackage{latexsym}
\usepackage[latin1]{inputenc}
\usepackage{graphicx}
\usepackage{amsmath}
\usepackage{palatino}
\usepackage{mathpazo}
\usepackage[normalem]{ulem}

\usepackage{booktabs}
\usepackage{dcolumn}

\def\jnl@style{\it}
\def\aaref@jnl#1{{\jnl@style#1}}

\def\aaref@jnl#1{{\jnl@style#1}}

\def\aj{\aaref@jnl{AJ}}                   % Astronomical Journal
\def\apj{\aaref@jnl{ApJ}}                 % Astrophysical Journal
\def\apjl{\aaref@jnl{ApJ}}                % Astrophysical Journal, Letters
\def\apjs{\aaref@jnl{ApJS}}               % Astrophysical Journal, Supplement
\def\apss{\aaref@jnl{Ap\&SS}}             % Astrophysics and Space Science
\def\aap{\aaref@jnl{A\&A}}                % Astronomy and Astrophysics
\def\aapr{\aaref@jnl{A\&A~Rev.}}          % Astronomy and Astrophysics Reviews
\def\aaps{\aaref@jnl{A\&AS}}              % Astronomy and Astrophysics, Supplement
\def\mnras{\aaref@jnl{Mon.~Not.~Roy.~Astron.~Soc.}}             % Monthly Notices of the RAS
\def\prd{\aaref@jnl{Phys.~Rev.~D}}        % Physical Review D
\def\prc{\aaref@jnl{Phys.~Rev.~C}}  % Physical Review C
\def\prl{\aaref@jnl{Phys.~Rev.~Lett.}}    % Physical Review Letters
\def\qjras{\aaref@jnl{QJRAS}}             % Quarterly Journal of the RAS
\def\skytel{\aaref@jnl{S\&T}}             % Sky and Telescope
\def\ssr{\aaref@jnl{Space~Sci.~Rev.}}     % Space Science Reviews
\def\zap{\aaref@jnl{ZAp}}                 % Zeitschrift fuer Astrophysik
\def\nat{\aaref@jnl{Nature}}              % Nature
\def\aplett{\aaref@jnl{Astrophys.~Lett.}} % Astrophysics Letters
\def\apspr{\aaref@jnl{Astrophys.~Space~Phys.~Res.}} % Astrophysics Space Physics Research
\def\physrep{\aaref@jnl{Phys.~Rep.}}      % Physics Reports
\def\physscr{\aaref@jnl{Phys.~Scr}}       % Physica Scripta
\def\commat{\aaref@jnl{Comm.~Math.~Phys.}}              % Communications in Mathematical Physics
\def\science{\aaref@jnl{Science}}               % Science
\def\cqg{\aaref@jnl{Classical Quant.~Grav.}}            % Classical and Quantum Gravity
\def\jpcs{\aaref@jnl{JPCS}}                                     % Journal of Physics Conference Series
\def\ijmpd{\aaref@jnl{Int.~J.~Mod.~Phys.~D}}                    % International Journal of Modern Physics D
\def\grg{\aaref@jnl{Gen.~Relat.~Gravit.}}               % General Relativity and Gravitation
\def\rpp{\aaref@jnl{Rep.~Prog.~Phys.}}          % Reports on Progress in Physics
\def\npa{\aaref@jnl{Nucl.~Phys.~A}}        % Nuclear Physics A
\def\lrr{\aaref@jnl{Living Rev.~Rel.}}                   % Living reviews in relativity
\def\jcap{\aaref@jnl{J.~Cosmology Astropart.~Phys.}}    % Journal of cosmology and astroparticle physics
\def\rmp{\aaref@jnl{Rev.~Mod.~Phys.}}   %Reviews of modern physics

\allowdisplaybreaks[1]
\begin{document}

	\title{Multi-scalar Gauss-Bonnet gravity: scalarized black holes beyond spontaneous scalarization}

	\author{Kalin V. Staykov}
	\email{kstaykov@phys.uni-sofia.bg}
	\affiliation{Department of Theoretical Physics, Faculty of Physics, Sofia University, Sofia 1164, Bulgaria}
	
	\author{Daniela D. Doneva}
	\email{daniela.doneva@uni-tuebingen.de}
	\affiliation{Theoretical Astrophysics, Eberhard Karls University of T\"ubingen, T\"ubingen 72076, Germany}
	\affiliation{INRNE - Bulgarian Academy of Sciences, 1784  Sofia, Bulgaria}

	\begin{abstract}
	Recently, a new nonlinear mechanism for black hole scalarization, different from the standard spontaneous scalarization, was demonstrated to exist for scalar Gauss-Bonnet theories in which no tachyonic instabilities can occur. Thus Schwarzschild black hole is linearly stable but instead nonlinear instability can kick-in.	
	In the present paper we extend on this idea in the case of multi-scalar Gauss-Bonnet gravity with exponential coupling functions of third and fourth leading order in the scalar field. The main motivation comes from the fact that these theories admit hairy compact objects  with zero scalar charge, thus zero scalar-dipole radiation, that automatically evades the binary pulsar constraints on the theory parameters.  We demonstrate numerically the existence of scalarized black holes for both coupling functions and for all possible maximally symmetric scalar field target spaces. The thermodynamics and the stability of the obtained solution branches is also discussed.
	\end{abstract}

	\maketitle
	
	\section{Introduction}

Spontaneous scalarization is a mechanism which allows for a given extended scalar-tensor theory (STT) to agree with the General relativity (GR) predictions in the weak field regime and exhibit significant deviations for strong fields \cite{Damour1993, Damour1996}. For the first time this mechanism was proposed by Damour and Esposito-Farese in \cite{Damour1993}, and until recently it was studied mainly in the context of neutron stars in which case the source of scalarization is the presence of matter (although see \cite{STEFANOV2007,Stefanov2008,Doneva2010,Cardoso2013a}). Recently, a new kind of spontaneous scalarization mechanism for black holes in scalar Gauss-Bonnet (sGB) gravity was demonstrated \cite{Doneva2018,Silva2018} in which case the source of scalarization is the curvature of the space-time itself (the so-called curvature induced scalarization). This mechanism for scalarization is active for neutron stars in sGB gravity \cite{Silva2018,Doneva2018a}. The curvature induced scalarization \cite{Doneva2018,Silva2018,Antoniou2018,Antoniou2018a,Doneva:2018rou,Minamitsuji2019,Silva2019,Brihaye2019,Myung2019,Hod2019,Cunha:2019dwb,Collodel:2019kkx} and  scalarization induces by rotation (the spin induced scalarization) \cite{Dima:2020yac,Hod:2020jjy,Doneva:2020nbb,Herdeiro:2020wei,Berti:2020kgk,Doneva:2020kfv} were thoroughly studied in multiple papers.  The phenomenon of curvature induced scalrization was demonstrated also for black holes \cite{Doneva2020} and neutron stars \cite{Staykov2022} in multi-scalar Gauss-Bonnet (MSGB) gravity as well.   

In the case of the above mentioned standard spontaneous scalarization, Schwarzschild black hole  is a solution of the field equations of the corresponding scalar-tensor theory for zero scalar field, and the scalarized branch bifurcates from it at some point, at which the GR solution becomes linearly unstable. In the majority of the studied cases there is a continuous transition between the stable nonscalarized and the stable scalarized black hole solutions at the bifurcation point. For the MSGB gravity, however, it was shown \cite{Doneva2020} that depending on the coupling function, there could be a jump between the stable Schwarzschild solution and the stable scalarized solutions and both are connected with an unstable scalarized black holes branch that can have some interesting astrophysical implications (see e.g. \cite{Doneva:2022byd}).    

In \cite{Doneva2021} the authors demonstrate the existence of new nonlinear mechanism for scalarization in a class of $Z_2$ symmetric  Gauss-Bonnet theories which does not allow for tachyonic instabilities to occur, thus Schwarzschild solution is always linearly stable. However, it is unstable against nonlinear perturbations which leads to a development of a static nonzero scalar field configuration. This is demonstrated by evolving the nonlinear equation for the scalar field and observing the development of the nontrivial scalar hair when the amplitude of the perturbation is large enough. The obtained scalarized black hole solutions form stable and unstable branches that are not continuously connected to the Schwarzschild one. The radial stability of these hairy black hole solutions was thoroughly investigated in \cite{BlazquezSalcedo2022}. In this paper the authors demonstrate that for any given value of the free parameter in the theory there is only one  radially sable scalarized branch -- either one starting from zero mass and terminated at maximal mass (for large values of the free parameter $\kappa$ in their coupling function) or one limited between two finite masses (for small $\kappa$). In addition, parts of the stable branches can lose hyperbolicity. Similar black hole solutions were found also in another theory of gravity, namely Einstein-Maxwell-scalar gravity \cite{Blazquez-Salcedo:2020nhs,LuisBlazquez-Salcedo:2020rqp}.

Some of the main restrictions on the scalar-tensor theories come from the absence of scalar dipole radiation from the observed binary pulsars and recently -- from the binary merger observations. Indeed, strong constraints have already been put on the sGB theory parameters through these mechanisms \cite{Danchev2021}.  Theories leading to hairy compact object solutions with zero scalar charge, however, can evade those constraints, which may allow for significantly wider ranges of the allowed values of the free parameters in those theories, hence larger deviations from GR. One particular class of such theories are the multi-scalar Gauss-Bonnet (MSGB) theories \cite{Doneva2020,Staykov2022}. In that case, the scalar field has leading order asymptotic $1/r^2$, hence the scalar charge is zero and so does the scalar dipole radiation. A parallel can be made with the findings in \cite{Ventagli:2021ubn} where the binary pulsar constraints can be evaded by extending the original sGB gravity to include an additional coupling between the Ricci scalar and the scalar field. 

This motivates us to extend on the results in \cite{Doneva2021} and explore  a similar scalarization mechanism in the case of multi-scalar Gauss-Bonnet gravity. In our case, however, we use both even and odd, in the scalar field, coupling functions, namely exponential functions with leading order $\chi^4$ and $\chi^3$. Note, that until now a coupling function $\chi^3$ is not considered even in the sGB case. We demonstrate the existence of static and spherically symmetric black hole solutions by directly solving the field equations for all possible maximally symmetric scalar field target spaces, namely spherical, hyperbolical and flat geometry. 

The paper is structured as follow. In Section II we briefly review the mathematical background of MSGB gravity and present the dimensionally reduced field equations. In Section III we present the numerical results for the two coupling functions we study. The paper ends with Conclusions.

	\section{Multi-scalar Gauss-Bonnet gravity }

    In this paper we study black hole scalarization in multi-scalar Gauss-Bonnet gravity with $N$ scalar fields $\varphi=(\varphi^1,...,\varphi^N)$ which take values on a patch of a $N$-dimensional Reimannian manifold, called $target\ space$, equipped with positively defined metric $\gamma_{ab}(\varphi)$. The reader, interested in more mathematical details, we refer to \cite{Damour1992,Doneva2020}.
	The most general form of the theory is defined by the following action
	\begin{eqnarray}
	S=&\frac{1}{16\pi G}\int d^4x \sqrt{-g} 
	\Big[R -  2g^{\mu\nu}\gamma_{ab}(\varphi)\nabla_{\mu}\varphi^{a}\nabla_{\nu}\varphi^{b} - V(\varphi) 
	+ \lambda^2 f(\varphi){\cal R}^2_{GB} \Big],\label{eq:quadratic}
	\end{eqnarray}
	where $R$ is the Ricci scalar with respect to the spacetime metric $g_{\mu\nu}$, and  $V(\varphi)$ is the potential of the scalar fields. The coupling function  $f(\varphi)$ depends only on $\varphi$, $\lambda$ is the Gauss-Bonnet coupling constant having  dimension of $length$ and ${\cal R}^2_{GB}$ is the Gauss-Bonnet invariant defined by ${\cal R}^2_{GB}=R^2 - 4 R_{\mu\nu} R^{\mu\nu} + R_{\mu\nu\alpha\beta}R^{\mu\nu\alpha\beta}$, where $R$ is the Ricci scalar, $R_{\mu\nu}$ is the Ricci tensor and $R_{\mu\nu\alpha\beta}$ is the Riemann tensor.  For the purpose of this study, we chose the case of theory with vanishing potential $V(\chi) = 0$.% and the coupling function will be specified later. 

    In order to study the problem we chose the target space to be 3-dimensional maximally symmetric space, namely $\mathbb{S}^3$, $\mathbb{H}^3$ or $\mathbb{R}^3$
    	with the metric 
    	\begin{eqnarray}
    	\gamma_{ab}(\varphi)d\varphi^a d\varphi^b= a^2\left[d\chi^2 + H^2(\chi)(d\varTheta^2 + \sin^2\varTheta d\Phi^2) \right],
    	\end{eqnarray}
    	where $a>0$ is a constant  and $\varTheta$ and $\Phi$ are the standard angular coordinates on the 2-dimensional sphere $\mathbb{S}^2$. The three possibilities for the target space are given by the metric function $H(\chi)$:  spherical $\mathbb{S}^3$, hyperbolic $\mathbb{H}^3$ and flat $\mathbb{R}^3$  geometry for $H(\chi)=\sin\chi$, $H(\chi)=\sinh\chi$, and $H(\chi)=\chi$ respectively. The parameter $a$ is related to the curvature $\kappa$ of the target space, where $\kappa=-1/a^2$ for  $\mathbb{S}^3$ and  $\kappa=1/a^2$ for $\mathbb{H}^3$. The scalar fields we chose in the following nontrivial way -- only $\chi = \chi(r)$ depends on the radial coordinate $r$. The other scalar fields do not depend on $r$ and they are given by $\varTheta = \theta$ and $\Phi = \phi$. The coupling function $f(\varphi)$ we take to depend on $\chi$ only. In this way the equations for $\varTheta$ and $\Phi$ separate from the rest.

	In the present paper we are interested in the static and spherically symmetric black hole solutions, hence we adopt the standard metric ansatz
	\begin{eqnarray}
	ds^2= - e^{2\Gamma}dt^2 + e^{2\Lambda}dr^2 + r^2(d\theta^2  + \sin^2\theta d\phi^2),
	\end{eqnarray} 
	where $\Gamma$ and $\Lambda$ depend on the radial coordinate $r$ only.

	 With the ansatz  for the scalar fields and using the above form of the metric, we obtain the following  reduced field equations 
	\begin{eqnarray}
	&&\frac{2}{r}\left[1 +  \frac{2}{r} (1-3e^{-2\Lambda})  \Psi_{r}  \right]  \frac{d\Lambda}{dr} + \frac{(e^{2\Lambda}-1)}{r^2} 
	- \frac{4}{r^2}(1-e^{-2\Lambda}) \frac{d\Psi_{r}}{dr} \nonumber \\ 
	&& \hspace{0.5cm} - a^2\left[ \left( \frac{d\chi}{dr}\right)^2 + 2e^{2\Lambda}\frac{H^2(\chi)}{r^2}\right] = 0, \label{DRFE1}\\ && \nonumber \\
	&&\frac{2}{r}\left[1 +  \frac{2}{r} (1-3e^{-2\Lambda})  \Psi_{r}  \right]  \frac{d\Gamma}{dr} - \frac{(e^{2\Lambda}-1)}{r^2} - a^2\left[ \left( \frac{d\chi}{dr}\right)^2 - 2e^{2\Lambda}\frac{H^2(\chi)}{r^2}\right] = 0,\label{DRFE2}\\ && \nonumber \\
	&& \frac{d^2\Gamma}{dr^2} + \left(\frac{d\Gamma}{dr} + \frac{1}{r}\right)\left(\frac{d\Gamma}{dr} - \frac{d\Lambda}{dr}\right)  + \frac{4e^{-2\Lambda}}{r}\left[3\frac{d\Gamma}{dr}\frac{d\Lambda}{dr} - \frac{d^2\Gamma}{dr^2} - \left(\frac{d\Gamma}{dr}\right)^2 \right]\Psi_{r} 
	\nonumber \\ 
	&& \hspace{0.5cm} - \frac{4e^{-2\Lambda}}{r}\frac{d\Gamma}{dr} \frac{d\Psi_r}{dr} + a^2\left(\frac{d\chi}{dr}\right)^2 =0, \label{DRFE3}\\ && \nonumber \\
	&& \frac{d^2\chi}{dr^2}  + \left(\frac{d\Gamma}{dr} \nonumber - \frac{d\Lambda}{dr} + \frac{2}{r}\right)\frac{d\chi}{dr} - \frac{2\lambda^2}{a^2r^2} \frac{df(\chi)}{d\chi}\left\{(1-e^{-2\Lambda})\left[\frac{d^2\Gamma}{dr^2} + \frac{d\Gamma}{dr} \left(\frac{d\Gamma}{dr} - \frac{d\Lambda}{dr}\right)\right]   \right. \nonumber \\
	&& \left. \hspace{0.5cm}  + 2e^{-2\Lambda}\frac{d\Gamma}{dr} \frac{d\Lambda}{dr}\right\} =  \frac{2}{r^2} H(\chi)\frac{dH(\chi)}{d\chi}e^{2\Lambda} \label{DRFE4}
	\end{eqnarray} 
	with 
	\begin{eqnarray}
	\Psi_{r}=\lambda^2 \frac{df(\chi)}{d\chi} \frac{d\chi}{dr}.
	\end{eqnarray}
	
	In order for equations (\ref{DRFE1})-(\ref{DRFE4}) to describe a black hole the following conditions at the horizon $r_H$, and at infinity should be satisfied: the existence of black hole horizon at $r=r_H$ requires 
	\begin{eqnarray}
	e^{2\Gamma}|_{r\rightarrow r_H} \rightarrow 0, \;\; e^{-2\Lambda}|_{r\rightarrow r_H} \rightarrow 0, \label{eq:BH_rh}
	\end{eqnarray} 
	and the asymptotic flatness imposes  
	\begin{eqnarray}
	\Gamma|_{r\rightarrow\infty} \rightarrow 0, \;\;  \Lambda|_{r\rightarrow\infty} \rightarrow 0,\;\; \chi|_{r\rightarrow\infty} \rightarrow 0\;\;.   \label{eq:BH_inf}
	\end{eqnarray} 	

	The asymptotic behavior of the metric functions and the scalar field one derives from the linearized equations at infinity: 
	\begin{equation}
	\Lambda\approx \frac{M}{r} + O(1/r^2), \;\; \Gamma\approx  -\frac{M}{r} + O(1/r^2), \;\; \chi\sim \frac{1}{r^2} .
	\end{equation}
    For the functions $\Gamma$ and $\Lambda$ we have the usual asymptotics. As for the scalar field, the scalar charge is zero since the leading order asymptotic is $1/r^2$.
	
    At the end, one should expand the field equations around the black hole horizon in order to derive the initial value for the first derivative of the scalar field at the horizon $\left(\frac{d\chi}{dr}\right)_H$. This gives us the following quadratic equation for $\left(\frac{d\chi}{dr}\right)_H$ \cite{Doneva2020}
    
    \begin{eqnarray}
    	&&\left(4\lambda^2 \left(a^2H(\chi_H)^2 - \frac{1}{2}\right) \left(\frac{df(\chi_H)}{d\chi}\right) r_H^3 + 8H(\chi_H)\left(\frac{dH(\chi_H)}{d\chi}\right) \left(\frac{df(\chi_H)}{d\chi}\right)^2 \lambda^4 r_H \right) \left(\frac{d\chi}{dr}\right)_H^2  \nonumber\\
    	&&\quad\nonumber\\
    	&&+ \left(\left(2a^2H(\chi_H)^2 - 1\right) r_H^4 + 8H(\chi_H)\left(\frac{dH(\chi_H)}{d\chi}\right) \left(\frac{df(\chi_H)}{d\chi}\right) \lambda^2 r_H^2 \right.\nonumber\\
    	&&\quad\nonumber\\
    	&&+ \left. 16\lambda^4 a^2 \left(a^2H(\chi_H)^2 - \frac{1}{2}\right) \left(\frac{df}{d\chi}\right)_H^2 H(\chi_H)^2\right) \left(\frac{d\chi}{dr}\right)_H +  2H(\chi_H)\left(\frac{dH(\chi_H)}{d\chi}\right) r_H^3 \nonumber \\
    	&&\quad\nonumber\\
    	&& - \left(\frac{df(\chi_H)}{d\chi}\right)\lambda^2 \left(\left(2a^2H(\chi_H)^2 - 1\right)^2 - 2\left(2a^2H(\chi_H)^2 - 1\right)\right) = 0  \label{eq:initial_cond_dchidr}.
    \end{eqnarray}
    
    In order for Schwarzschild black hole to be a solution of the field equations one should chose the root with the positive sign. Requiring the solutions of the quadratic equation to be real, we obtain the following existence condition  	
\begin{eqnarray} \label{eq:exist}
&&a^2 \left( 2 a^2 H(\chi_H)^{2}-1 \right) ^{2} \left( 
64{H(\chi_H)}^{4}{a^2}{ \left(\frac{df}{d\chi}\right)}^{4}_H{\lambda}^{8}+32
{H(\chi_H)}^{2}{a^2}{\left(\frac{df}{d\chi}\right)}^{2}_H {\lambda}^{4}{r_H}^
{4}\right.  \nonumber \\ &&\left.+96{H(\chi_H)}{\left(\frac{dH(\chi_H)}{d\chi}\right)}{\left(\frac{df}{d\chi}\right)}^{3}_H {\lambda}^{6}{r_H}^{2}-24{\left(\frac{df}{d\chi}\right)}^{2}{\lambda}^{4}{r_H}^{4}+{ a^2}{r_H}^{8} \right) \geq 0.
\end{eqnarray}

\section{Numerical setup and results}	

As we mentioned, we study MSGB theories which does not exhibit tachyonic instabilities. The coupling function in this case should satisfy the following conditions \cite{Doneva2021}
\begin{eqnarray}
f(0) =1, \quad \frac{df}{d\chi}(0) = 0, \quad \frac{d^2f}{d\chi^2}(0) = 0.
\end{eqnarray}

The first one is a normalization condition that can be assumed without loss of generality. The second one ensures that Schwarzschild black hole is a solution of the field equations for $\chi = 0$. The third condition leads to the fact that there are no tachyonic instabilities, and the Schwarzschild black hole is stable against linear perturbations.

In this section we present the numerical results for two exponential coupling functions which satisfy the above conditions, namely

\begin{equation}
f_1(\chi) = \frac{1}{4\beta}\left(1-e^{-\beta\chi^4}\right), \label{eq:f1}
\end{equation}
and
\begin{equation}
f_2(\chi) = \frac{1}{3\beta}\left(1-e^{-\beta\chi^3}\right), \label{eq:f2}
\end{equation}
where $\beta$ is a parameter. For both coupling functions we will present results for $\beta = 0.5$ but wider range of values were studied, and the effect of $\beta$ will be mentioned in the text.

The first coupling function $f_1(\chi)$ is of leading order $\chi^4$, and the second one $f_2(\chi)$ -- of $\chi^3$. We have studied coupling function in the form $f(\chi) = \chi^4$, and  $f(\chi) = \chi^3$ as well, however, in those cases, we were not able to find any stable black hole solutions. 

In the results presented below the Gauss-Bonnet coupling constant $\lambda$ in the field equations (\ref{DRFE1})-(\ref{DRFE4}) is used as a normalization parameter.

\subsection{Results for coupling function $f_1(\chi)$}

In this section we present the results for the coupling function $f_1(\chi)$ (\ref{eq:f1}). In this case the coupling function and the field equations are symmetric with respect to the sign of the scalar field, and solutions with both signs exist. Here we present only the results for $\chi > 0$ without loss of generality. In addition, even if we fix the theory parameters $a^2$, $\lambda$ and $\beta$, multiple branches of solutions exist which can be labeled by the number of the nodes of the scalar field. We present only the results for the branches with no nodes of the scalar field (the fundamental branch) since all the rest  are presumably unstable \cite{BlazquezSalcedo2018}.

\begin{figure}[]
	\centering
	\includegraphics[width=0.45\textwidth]{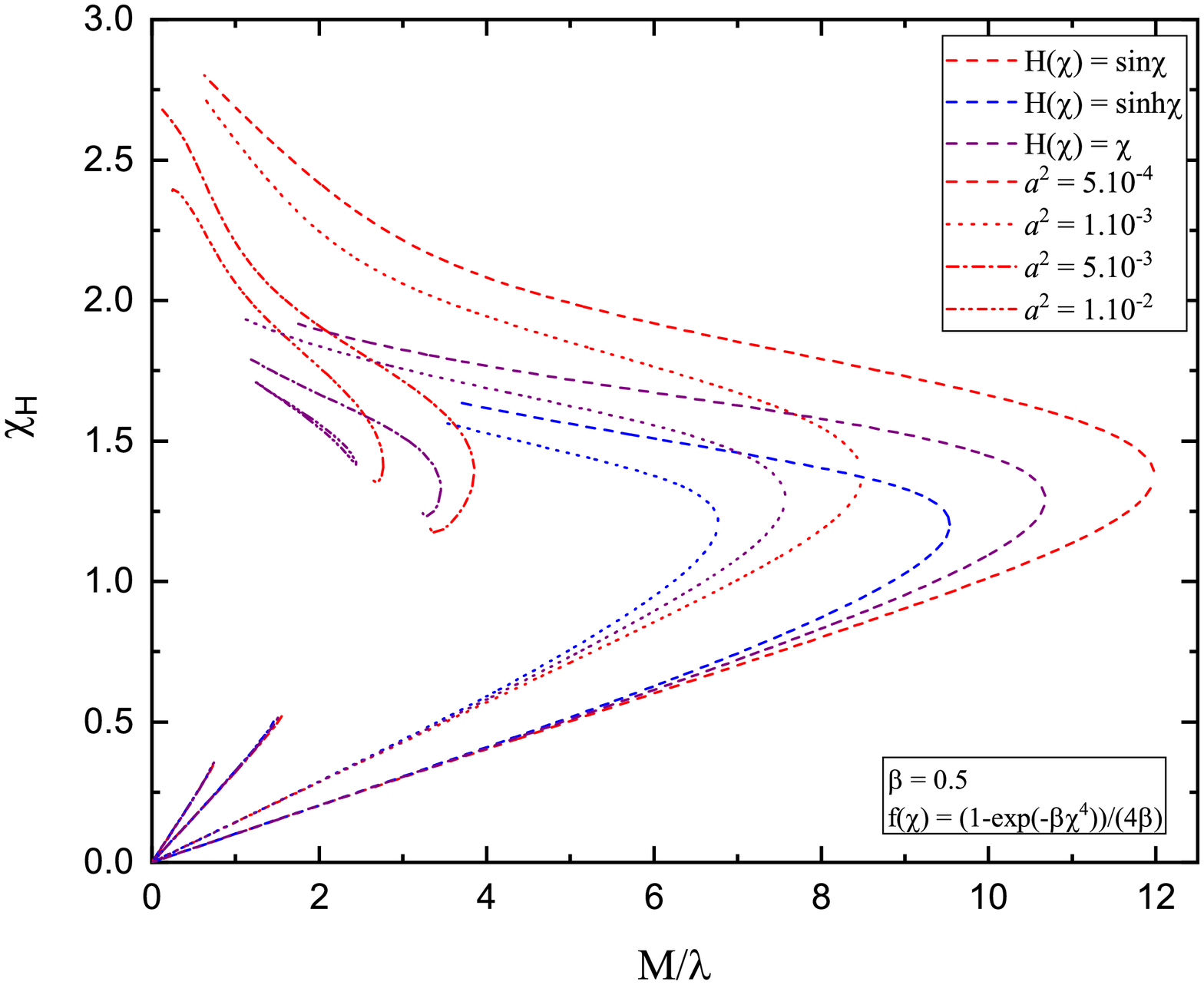}
	\includegraphics[width=0.47\textwidth]{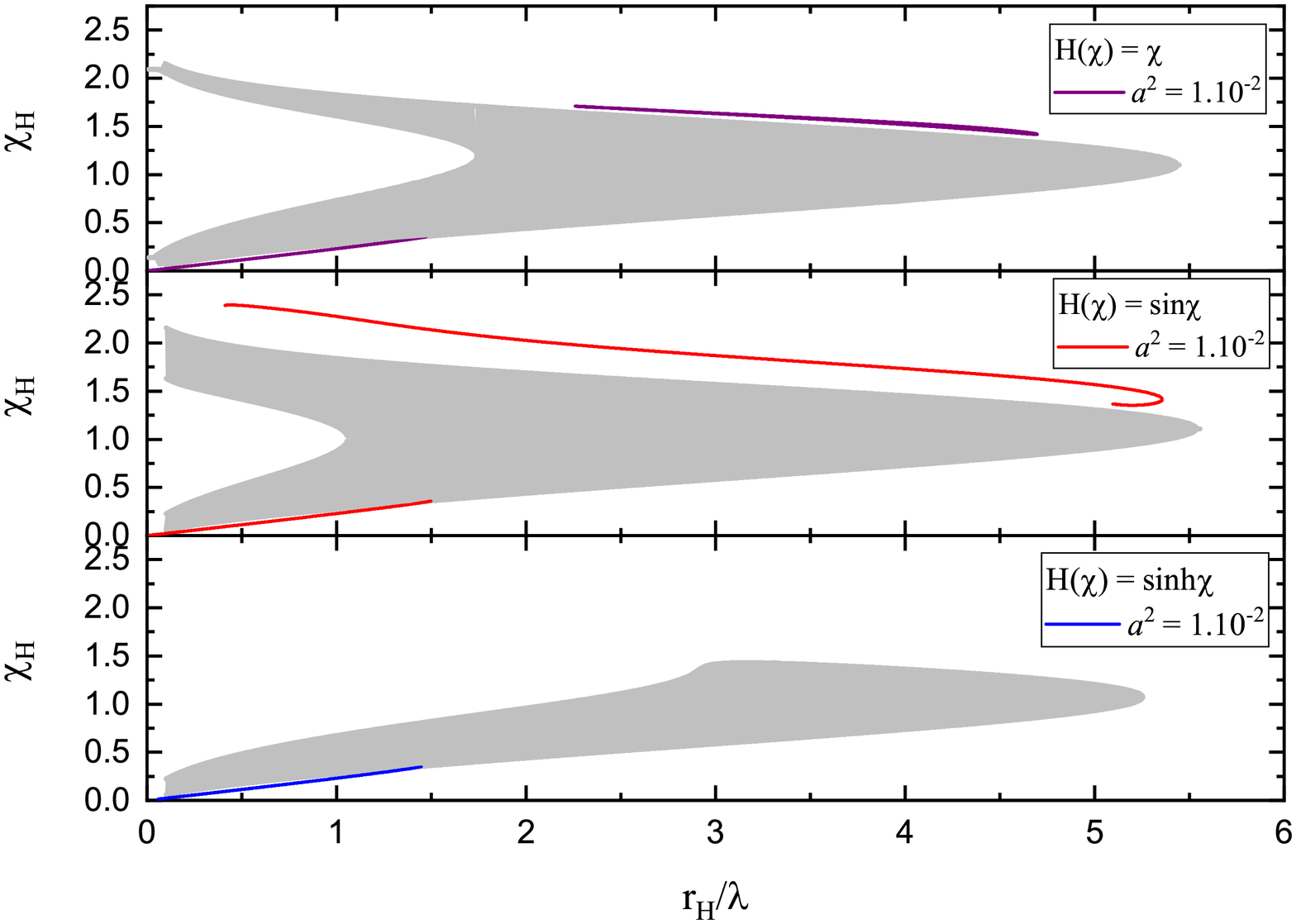}
	\caption{Left: Scalar field on the horizon as a function of the black hole mass. Right: The restricted from the existence condition \eqref{eq:exist} area, and the corresponding branches of solutions for all three cases for $H(\chi)$ and one value for $a^2$. }
	\label{Fig:chi_chi4}
\end{figure}

In the left panel of Fig. \ref{Fig:chi_chi4} the scalar field on the horizon as a function of the mass of the black hole is plotted. The presented results are for a fixed value of $\beta$, all possible cases of $H(\chi)$ and a wide range for $a^2$. In all cases there is a scalarized branch which starts from the origin (zero mass and zero scalar field), and the scalar field on the horizon increases with the increase of the mass of the black hole. For the smaller values of $a^2$, this branch reaches a maximal mass at which it turns around, forming a new branch for which the scalar field continues to increase as the mass decreases. Those upper branches tend to non zero scalar field at zero mass, however, they get terminated at some small value of the black hole mass due to numerical difficulties connected to the increasing stiffness of the field equations. 

For intermediate values of $a^2$ the lower branch that starts at the origin is terminated due to violations of the regularity  condition (\ref{eq:exist}) before the maximal mass is reached. In this case, however,  an upper branch  that tends to a nonzero $\chi_H$ when $M\rightarrow 0$ still exists. In this case the scalar field  decreases with the increase of the black hole mass until a maximum mass is reached and the branch  turns left. This second part of the branch is short and it ends with what looks like an inspiraling part.  

For small and intermediate $a^2$ the observed behavior of the solutions was qualitative the same for all choices of $H(\chi)$.  If we increase $a$ further (in our case $a^2 = 1\times10^{-2}$) we can observe a  qualitatively different behavior between the three function $H(\chi)$. For all $H(\chi)$   the lower branch behaves the same -- it starts at the origin and it is terminated due to violations of the regularity  condition (\ref{eq:exist}) before the maximal mass is reached. What differs is the upped branches. For $H(\chi)=\sin{\chi}$ these branches are the same as the intermediate $a^2$ -- after reaching a maximum mass the upper branch starts inspiraling. Interestingly,  for $H(\chi)=\sinh{\chi}$ we could not find an upped branch of solutions despite our best effort. It is not possible, though, to rigorously assess  whether this is a numerical problem or true disappearance of upper branch solutions. The most interesting is the $H(\chi)=\chi$ case where  two more branches exist which does not start from zero scalar field. Instead, they are connected both at the maximal and at the minimal mass. Thus, in this case  the upper branches do not exist for arbitrary small $M/\lambda$.

In order to indeed prove that the observed termination of the branches is due to regularity violation (see eq. (\ref{eq:exist})) and is not a numerical artifact, we performed the following analysis.  In the right panel of Fig. \ref{Fig:chi_chi4}, we present the restricted region of the parameter space where condition (\ref{eq:exist}) is violated and thus no hairy black hole solutions can exist. All plots are for $a^2 = 1\times10^{-2}$ but different functions $H(\chi)$. In each figure, the sequences of hairy black hole solutions for this same $a^2$ and $H(\chi)$ are also plotted.  Let us focus on $H(\chi) = \sinh(\chi)$ (bottom plot of the right panel in Fig. \ref{Fig:chi_chi4}). In that case we were able to find only one lower branch of solutions (blue line) that terminates as it crosses the restricted gray ares. One can see that this area has  qualitatively different shape for $H(\chi) = \sinh(\chi)$, compared to the other two plots. It is interesting to note that this is the case for which we were not able to find upped solution branch starting from non zero mass and scalar field. As for the other two functions $H(\chi) = \chi$ and $H(\chi) = \sin(\chi)$ (upped and middle panel) the lower branch gets again terminated as it crosses the gray area, while the upper branch never reaches it because in the  high mass region of this branch either an inspiraling sequence of solutions is present (middle panel) similar to the single scalar field case \cite{BlazquezSalcedo2022} or two connected branches exist forming something closely resembling a loop (upper panel). 

The right panel of  Fig.  \ref{Fig:chi_chi4} is for a fixed value of $a^2$. Our calculations show, though, that when the value of $a^2$ decreases sufficiently (in the figure $a^2=10^{-3}$ and lower) all three cases get similar bell shaped restricted area, and the corresponding branches of solutions are not affected by the existence condition. That is why this case is not explicitely demonstrated here. Concerning the effect the parameter $\beta$ has on the results -- we have studied values of $\beta$ spanning multiples orders of magnitude. The general effect of increasing $\beta$ is that it lowers the maximal black hole mass the sequences reach and lowers the values of the scalar field on the horizon for same-mass black holes.

At this point we should compare our results with those in \cite{Doneva2021} for sGB with a single scalar field. Our coupling function $f_1$ is the same as the coupling function $f_1$ (eq. (3)) in   \cite{Doneva2021}. The only difference is connected to the notations -- the parameter $\kappa$ in  \cite{Doneva2021} is named $\beta$ in our case.  In both cases there are up to three branches of scalarized solutions and the lower branches which start from $M=0$ always exist. While in the sGB case the qualitative behavior of the branches is determined by the parameter $\kappa$, in the MSGB case a major role plays also the parameter $a^2$ connected to the curvature of the target space metric. We have also an additional freedom that is the choice of $H(\chi)$.  Still different choices of $a^2$ and $H(\chi)$ do not seem to lead to qualitatively new types of branches compares to sGB gravity but instead one can observe (sometime significant) quantitative differences. Something else important to note is that for some ranges of parameters we were not able to find upped branch of black hole solutions in MSGB gravity. We were not able to prove, though, whether this is a solid theoretical result or a numerical artifact.

Now, let us turn to the stability of the obtained black hole solutions. A proper analysis will require a study of their linear stability similar to \cite{BlazquezSalcedo2022} that is beyond the scope of the present paper. Let us remember, though, that the qualitative behavior of the branches in sGB and MSGB gravity is very similar. So one might guess that similar to sGB gravity, the lower branch of solutions is probably unstable while the upper one, having the largest scalar field for a given $M/\lambda$ among all existing branches, is stable (if exists).

\begin{figure}[]
	\centering
	\includegraphics[width=0.45\textwidth]{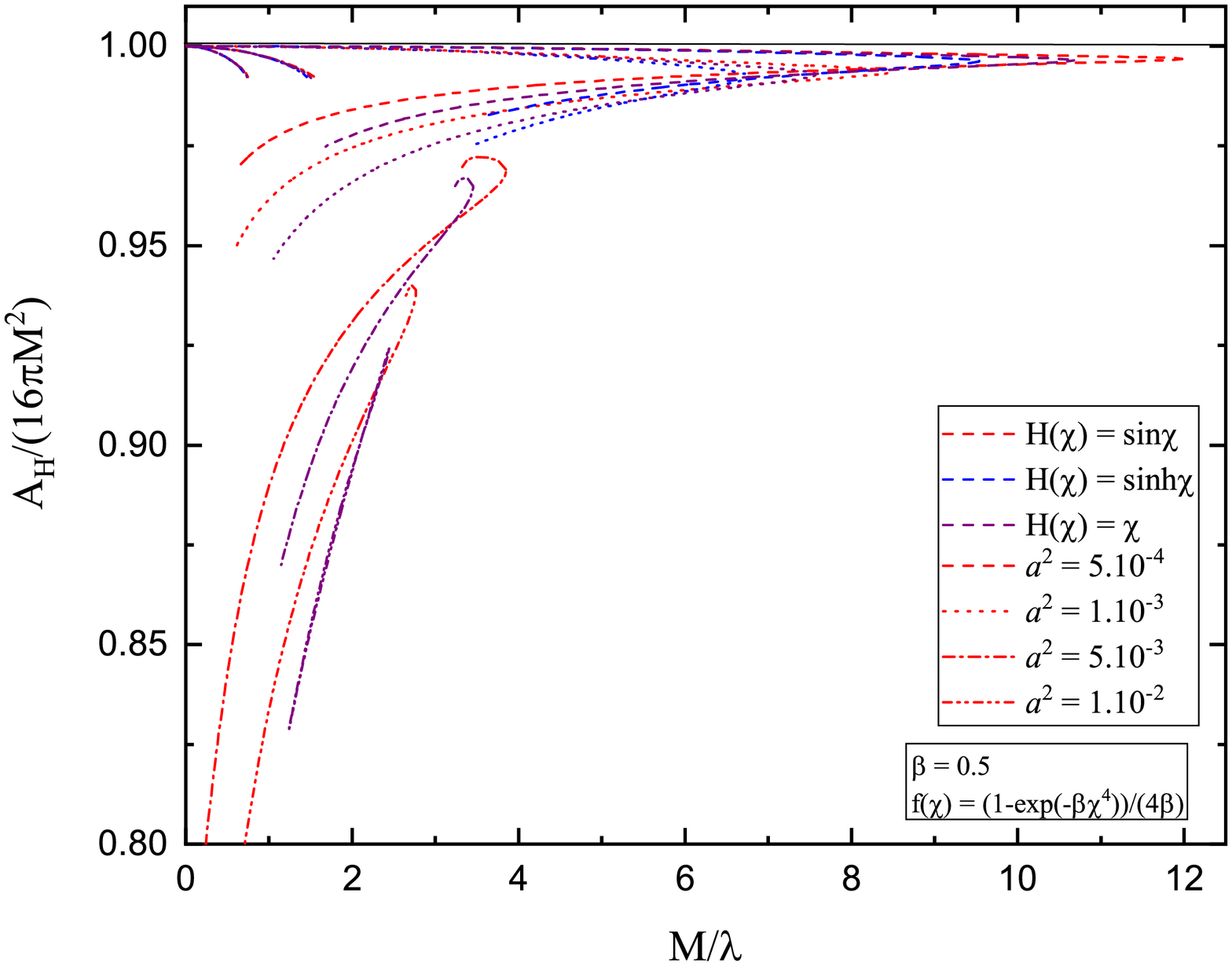}
	\includegraphics[width=0.45\textwidth]{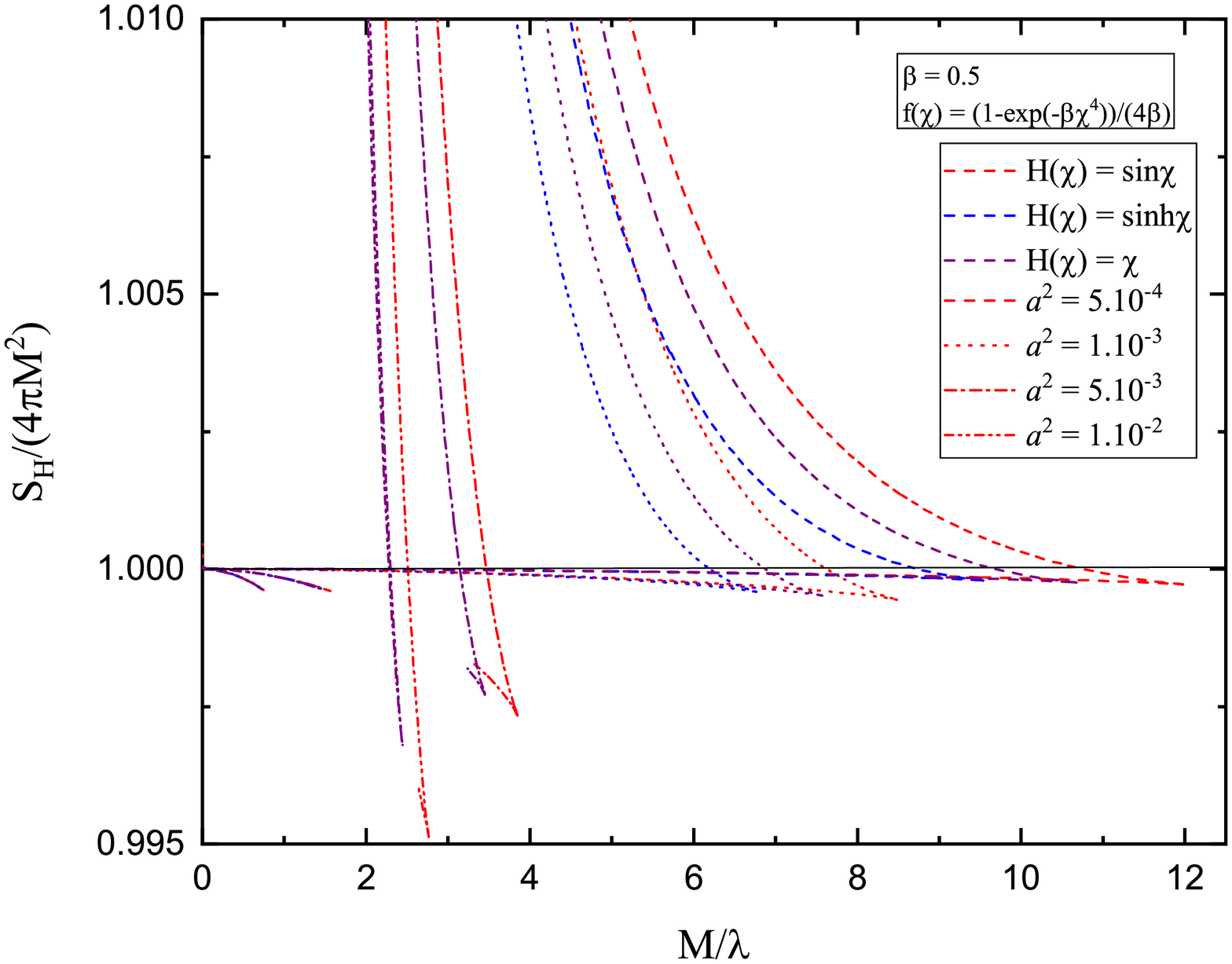}
	\caption{Left: The normalized to the Schwarzschild limit  area of the black hole horizon $A_H/16\pi M^2$ as a function of the black hole mass. Right: The normalized to the Schwarzschild limit  entropy $S_H/4\pi M^2$ as a function of the black hole mass. }
	\label{Fig:A_S_chi4}
	\end{figure}

 Similar conclusion can be drawn from examining the black hole horizon area $A_H$ and its entropy defined as
\begin{equation}
S_H = \frac{1}{4}A_H + 4\pi \lambda^2 f(\chi_H).
\end{equation}
In the left panel of Fig. \ref{Fig:A_S_chi4} we present the normalized to the Schwarzschild limit  area of the black hole horizon $A_H/16\pi M^2$ as a function of the black hole mass. In all cases the area of the scalarized solutions is smaller, compared to the Schwarzschild case. It is interesting to note that the upper  branches which start from the maximal mass have significantly smaller area, compared to the lower branches which start from $M/\lambda=0$ and $\chi_H=0$. In the right panel of Fig. \ref{Fig:A_S_chi4}  we present the normalized to the Schwarzschild limit  entropy $S_H/4\pi M^2$ as a function of the black hole mass. The weaker scalar field  branches  always have smaller entropy  compared to the Schwarzschild one that is another sign for their instability.  As for the upper branches --  the entropy is larger compared to Schwarzschild for most of the parameters space and only close to the maximum $M/\lambda$ an intersection between the two curves is observed. This indicates that at least part of the upper branch of solutions is energetically favorable over GR and thus probably stable.  The results for the black hole entropy are in correlation with our guess about the stability of the solutions we made above. Therefore, we expect that here, similar to  \cite{Doneva2021},  again there will be a jump between the stable scalarized black holes and the GR ones. As commented, this is contrary to the standard spontaneous scalarization where the  transition is smooth in the majority of the cases.

\subsection{Results for coupling function $f_2(\chi)$}

We continue our study of the nonlinear scalarization in MSGB gravity with the coupling function $f_2(\chi)$. In this case the coupling function and the field equations are not symmetric with respect to the change of sign of the scalar field, however, in the present study we concentrated only on the case $\chi > 0$.  We should note that scalarized solutions with such coupling were not obtained until now in sGB gravity.

\begin{figure}[]
	\centering
	\includegraphics[width=0.45\textwidth]{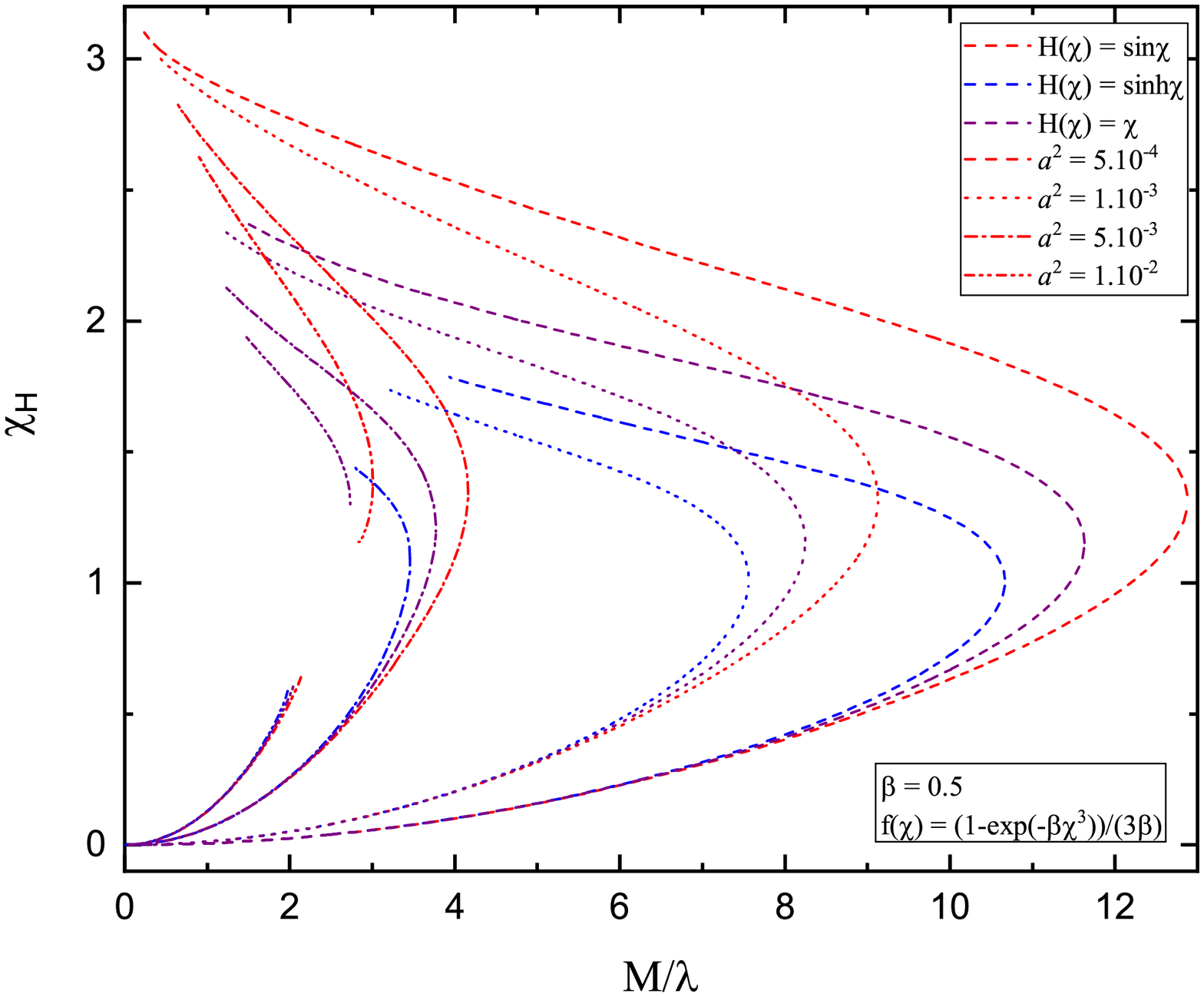}
	\includegraphics[width=0.47\textwidth]{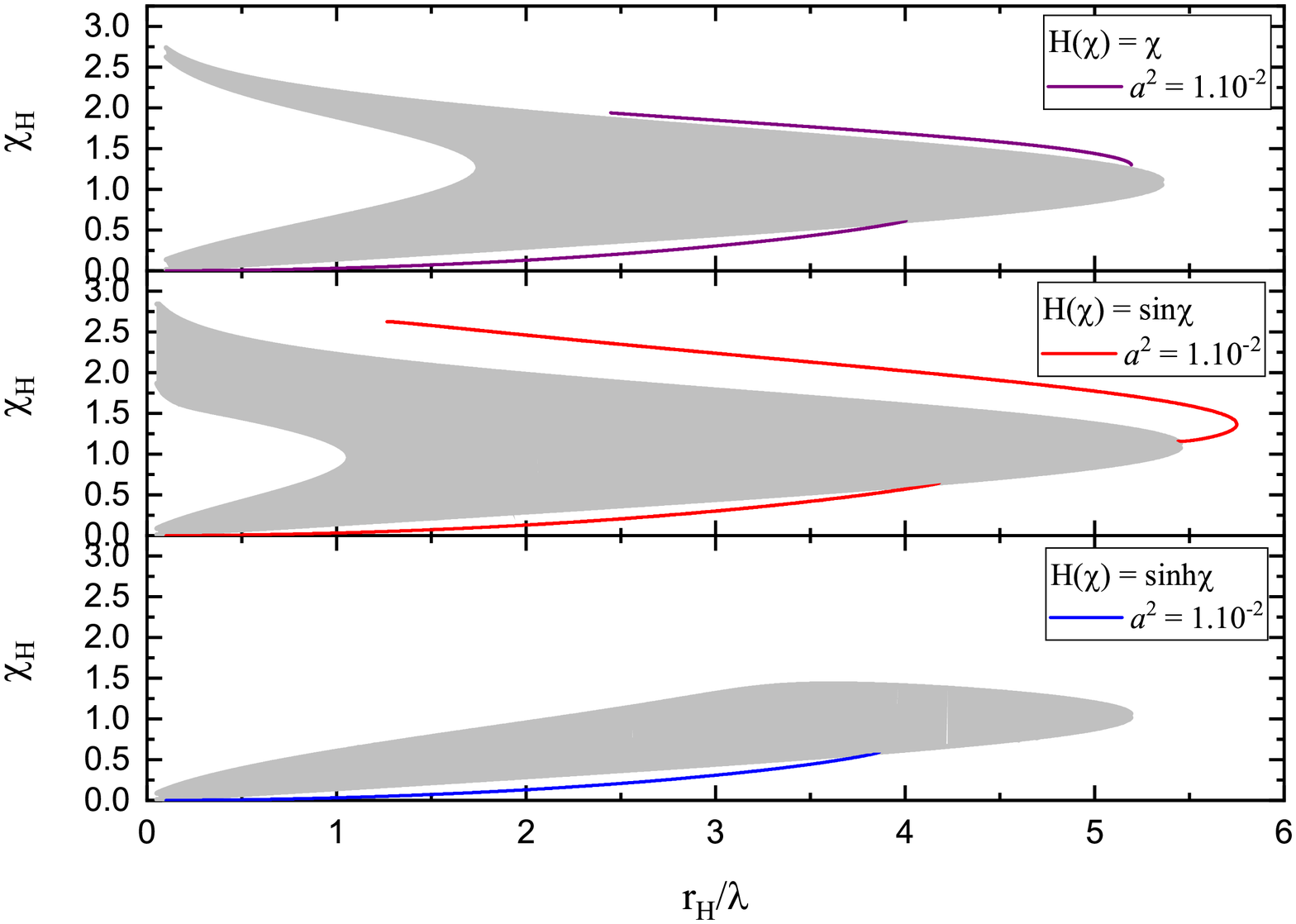}
	\caption{Left: Scalar field on the horizon as a function of the black hole mass. Right: The restricted from the existence condition \eqref{eq:exist} area, and the corresponding branches of solutions for all three cases for $H(\chi)$ and one value for $a^2$.}
	\label{Fig:chi_chi3}
\end{figure}

In the left panel of Fig. \ref{Fig:chi_chi3} we present the scalar field on the black hole horizon as a function of its mass. The  results are for all three possible cases for $H(\chi)$, and different values of the parameter $a^2$ (the same as in the previous case). For all cases, the observed behavior is qualitatively the same as for the coupling function $f_1$, with the main difference being that the branches which start at the origin, and get terminated before the maximal mass (due to regularity condition violation) start appearing for higher values of $a^2$. 
In the right panel we present the forbidden from the existence condition \eqref{eq:exist} area, and the corresponding solution branches as a function of the radius of the horizon for $a^2 = 1\times10^{-2}$. Similarly  to the $f_1({\chi})$ case, the restricted area for $H(\chi) = \sinh(\chi)$ has qualitatively different shape, compared to the other two cases, and we cannot find an upper branch of solutions above the restricted area contrary to the case of smaller $a^2$ and other $H(\chi)$.

\begin{figure}[]
	\centering
	\includegraphics[width=0.45\textwidth]{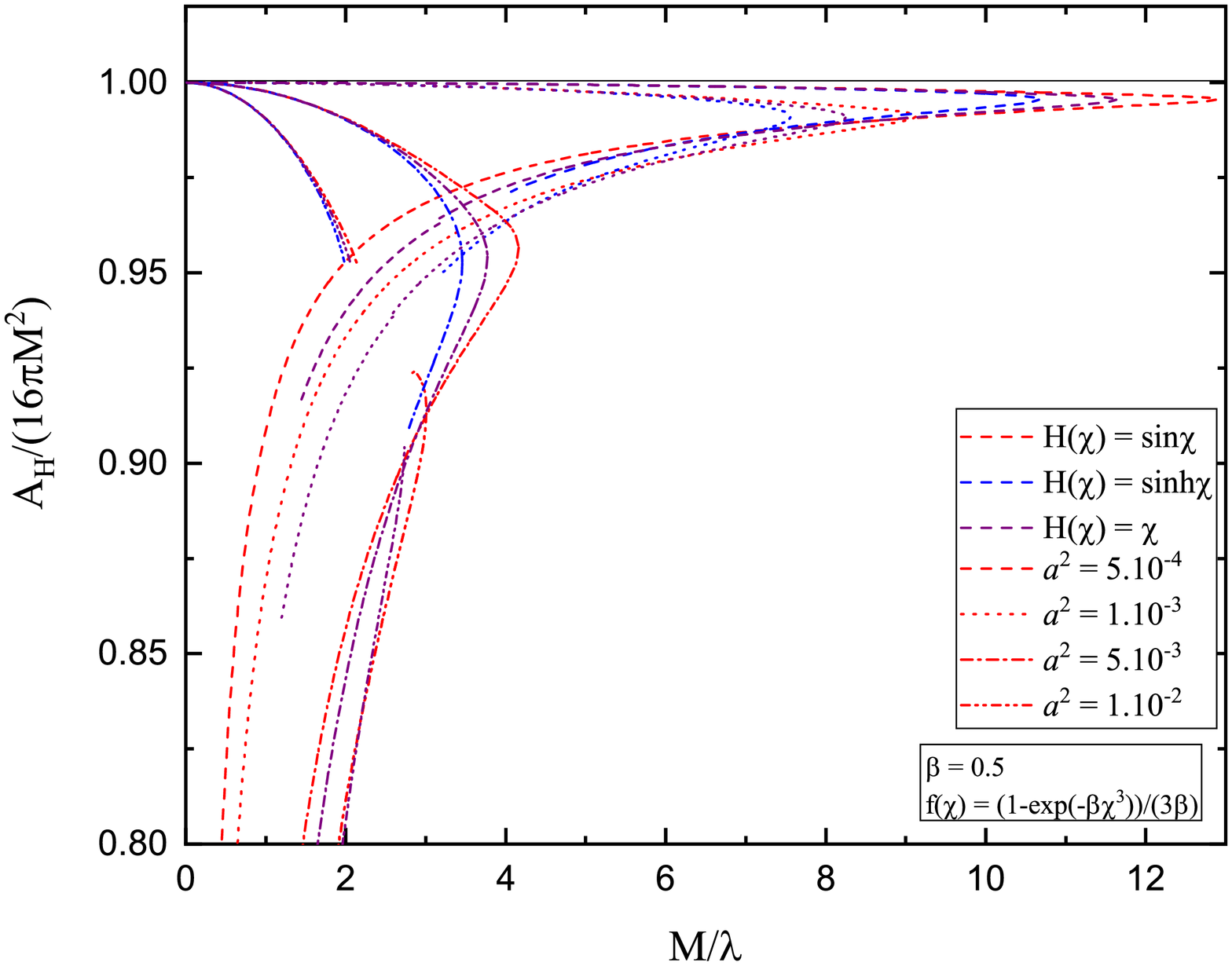}
	\includegraphics[width=0.45\textwidth]{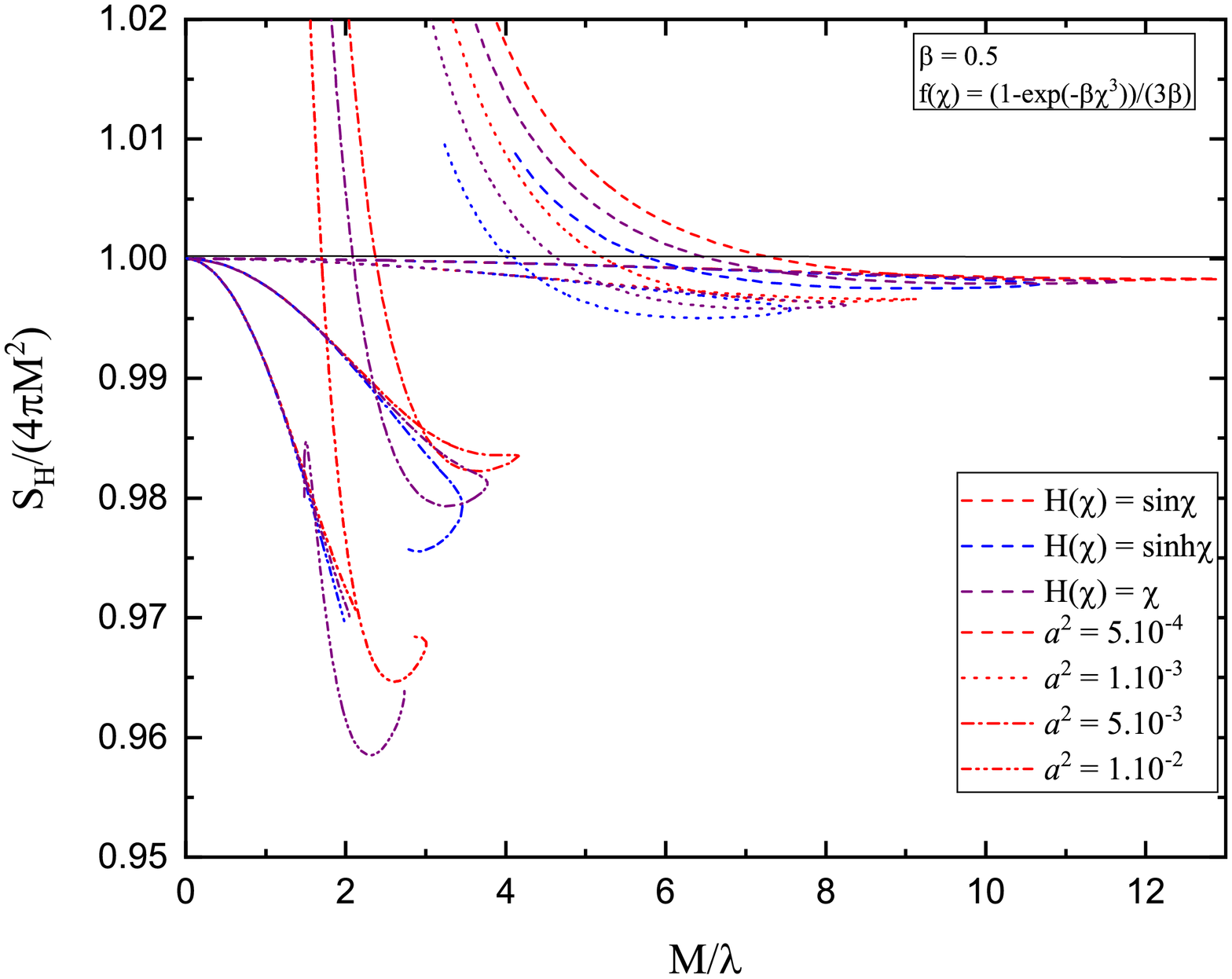}
	\caption{Left: The normalized to the Schwarzschild limit  area of the black hole horizon $A_H/16\pi M^2$ as a function of the black hole mass. Right: The normalized to the Schwarzschild limit  entropy $S_H/4\pi M^2$ as a function of the black hole mass.  }
	\label{Fig:A_S_chi3}
\end{figure}

We continue our study of the scalarized solutions with coupling function $f_2(\chi)$ with the area of the black hole horizon $A_H$, and the black hole entropy $S_H$.
In the left panel of Fig. \ref{Fig:A_S_chi3} we plot the normalized area of the black hole horizon as a function of the black hole mass. The behavior is qualitatively the same as the one for $f_1(\chi)$. However, for small values of $a^2$, BHs with small masses have significantly smaller area.  In the right panel we plot the normalized to the Schwarzschild limit entropy $S_H/4\pi M^2$ as a function of the black hole mass. In this cases as well, the behavior is very similar with the $f_1(\chi)$ case and only portions of the upper branches can have entropy larger than the Schwarzschild one making them thermodynamically preferred and probably stable.  

\section{Conclusion}

In the present paper we demonstrated the existence of the recently discovered mechanism for nonlinear scalarization in multi-scalar Gauss-Bonnet (MSGB) gravity, extending previous results in the single scalar field theory \cite{Doneva2021}. The new mechanism is different from the standard spontaneous scalarization, and it allows for the nonlinear formation of scalarized black holes, while at the same time the Schwarzschild solution is always linearly stable and thus no bifurcations occur.  The interesting fact about these solutions is that their scalar charge is zero leading to suppression of the scalar dipole radiation in dynamical processes. Thus, such a theory can easily evade strong observational constraints coming e.g. by the binary pulsar observations.

In the present work we studied two exponential coupling functions -- one of leading order $\chi^4$ in the scalar field and a new one, compared to previous single scalar field studies, of leading order $\chi^3$. Both coupling functions admit  Schwarzschild black hole as a solution of the field equations, and does not allow for tachyonic instabilities (the second derivative of the coupling function is zero for vanishing scalar field), hence no standard spontaneous scalarization can be observed.

We studied all three cases of maximally symmetric target spaces, namely spherical $H(\chi)= \sin(\chi)$, flat $H(\chi)= \chi$, and hyperbolic $H(\chi)= \sinh(\chi)$, a wide range of values for the parameter $a^2$ in the target space metric, and a wide range for the parameter $\beta$ in the coupling function. 
For both coupling functions the general behavior is qualitatively similar. For all value of $a^2$, $\beta$ and $H(\chi)$ a lower branch exists which starts from zero scalar field and zero mass. Along this branch the scalar field increases with the increase of the black hole mass  until either a maximal mass is reached (for smaller $a^2$) or the branch is terminated due to violation of the regularity condition (for larger $a^2$). The upper branch of solutions is characterized by an increasing scalar field as $M = 0$ is approached. At larger masses this branch either merges with the lower branch or inspirals after the maximum mass point is reached. For larger $a^2$ and some $H(\chi)$ two upper branches were observed that connect both at small and high masses, forming something resembling a closed loop. For larger $a^2$ and some $H(\chi)$,  we were not able to find  upper branches at all.

By making an analogy with the results in \cite{BlazquezSalcedo2022} for sGB with single scalar field, as well as examining the thermodynamics properties of the obtained solutions, the following conclusions for the stability of  black hole in the MSGB gravity can be drawn. We expect that in MSGB case the stable black hole branches are the once with maximum scalar field for a fixed black hole mass while all the rest of the branches are unstable.  Thus, just like in the sGB case, the stable black hole solutions are not continuously connected to the stable Schwarzschild branch. This my have significant footprint in the astrophysical observations if a transition between a scalarized and a nonscalarized phase occurs. This will happen with a jump during which all scalar hair will be radiated.

	\section*{Acknowledgements}
    KS acknowledges financial support by the Bulgarian NSF Grant KP-06-H28/7. DD acknowledge financial support via an Emmy Noether Research Group funded by the German Research
    Foundation (DFG) under grant no. DO 1771/1-1.  
    
	%%%%%%%%%%%%%%%%%%%%%%%%%%%%%%%%%%%%%%%%%%%%%%%%%%%%%%%%%%%%%%%%%%%%%%%%%%%%%%%
	
	\bibliography{references}

\end{document}